\documentclass[natbib]{svjour3}       
\smartqed  

\usepackage{graphicx}
\usepackage{bm,amssymb,amsmath,color,url,hyperref}
 \journalname{my journal}

\newcommand{\aap}{{Astron. Astrophys.}}

\newcommand{\apj}{{Astrophys. J.}}
\newcommand{\apjs}{{Astrophys. J. Suppl.}}
\newcommand{\apjl}{{Astrophys. J. Lett.}}
\newcommand{\nat}{{Natur.}}

\newcommand{\physrep}{{Phys. Rep.}}
\newcommand{\mnras}{{Mon. Not. Roy. Astron. Soc.}}

\newcommand{\araa}{{Annu. Rev. Astron. Astr.}}
\newcommand{\nar}{{New Atron. Rev.}}

\newcommand{\prd}{{Phys. Rev. D}}

\newcommand{\actaa}{{Acta Astronomy}}

\newcommand{\cjaa}{{Chin. J. Astron. Astr.}}

\newcommand{\beqa}{\begin{eqnarray}}
\newcommand{\eeqa}{\end{eqnarray}}
\newcommand{\be}{\begin{equation}}
\newcommand{\ee}{\end{equation}}
 \newcommand{\ba}{\begin{eqnarray}}
\newcommand{\ea}{\end{eqnarray}}

\begin{document}

\title{GRB Observational Properties
}


\author{Bing Zhang \and  Hou-Jun L\"{u} \and En-Wei Liang }


\institute{Bing Zhang  \at
              Department of Physics and Astronomy,
University of Nevada Las Vegas, NV 89154, USA\\
              \email{zhang@physics.unlv.edu}
           \and
Hou-Jun L\"{u} \at
              GXU-NAOC Center for
Astrophysics and Space Sciences, Department of Physics, Guangxi
University, Nanning 530004, China \\
Guangxi Key Laboratory for
Relativistic Astrophysics, Nanning, Guangxi 530004, China\\
              \email{lhj@gxu.edu.cn}           
              \and
En-Wei Liang \at
               GXU-NAOC Center for
Astrophysics and Space Sciences, Department of Physics, Guangxi
University, Nanning 530004, China \\
Guangxi Key Laboratory for
Relativistic Astrophysics, Nanning, Guangxi 530004, China\\
              \email{lew@gxu.edu.cn}
}

\date{Received: date / Accepted: date}

\maketitle

\begin{abstract}
We summarize basic observational properties of gamma-ray bursts (GRBs), including prompt emission properties, afterglow properties, and classification schemes. We also briefly comment on the current physical understanding of these properties.
\keywords{Gamma-ray bursts}
\end{abstract}

\section{Introduction}
\label{sec:intro}

Gamma-ray bursts (GRBs) are energetic bursts of $\gamma$-rays from deep space. They are the most luminous objects in the universe, signifying births of stellar-mass black holes or millisecond magnetars in most violent explosions, such as core collapse of rapidly spinning massive stars or mergers of two compact stars (NS-NS or NS-BH). Many in-depth reviews have been written on the subject: e.g. \citep{Fishman1995,Piran1999,vanParadijs2000,Meszaros2002,Zhang2004,Piran2004,Meszaros2006,Zhangb2007,Gehrels2009,Kumar2015}. This chapter summarizes the basic observational properties of GRBs, including prompt emission (\S\ref{sec:prompt}), afterglow (\S\ref{sec:afterglow}), and classification schemes (\S\ref{sec:classification}). Besides outlining the observational facts, a brief discussion on the current physical understanding of the data is also presented.

\section{Prompt emission}
\label{sec:prompt}
GRB prompt emission is usually defined as the soft $\gamma$-ray / hard X-ray emission detected by the GRB triggering detectors (such as {\em Swift}/BAT and {\em Fermi}/GBM). Some GRBs are detected in higher (e.g. GeV) or lower (e.g. optical) energies during the prompt emission phase.

\subsection{Temporal properties}

\subsubsection{Duration distribution}
GRB duration is usually quantified as $T_{90}$, the duration during which 5\% to 95\% of the total fluence is detected by the triggering detector. The definition of $T_{90}$ depends on the energy band and sensitivity of the detector. A same burst may have a longer $T_{90}$ if the detector is more sensitive or has a softer energy band.

In the CGRO/BATSE band (25-350 keV), GRBs showed a bimodal distribution with a rough separation at 2 s \citep{Kouveliotou1993}. This was the foundation of classifying GRBs into long ($T_{90} > 2$ s) and short ($T_{90} < 2$ s) categories (Fig.1a). The bimodal distribution was confirmed by other later detectors \citep[e.g.,][]{Sakamoto2011,Paciesas2012,Zhangfw2012,Lien2016}, even though the relative fractions of the two types vary for different detectors and also for different energy bands in the same detector \citep{Qin2013}. The bimodal classification is more evident in the two-dimensional $T_{90}-{\rm HR}$ domain, where HR is the hardness ratio of a burst. Long GRBs are typically softer than short GRBs \citep[][see Fig.1b]{Fishman1995}. Broad-band observations now show that the two duration classes roughly correspond to two types of progenitor systems: long GRBs are related to deaths of massive stars, and short GRBs are related to mergers of compact objects (see \S\ref{sec:classification} for more discussion).

\subsubsection{Lightcurves}

GRB lightcurves are erratic, usually display overlapping emission episodes with a wide range of variability time scales \citep{Fishman1995}. Some bursts have one or several clearly separated ``pulses'', which typically show a fast-rising and exponential-decay (FRED) shape. The duration of these pulses is typically seconds. Some GRBs have a ``precursor" followed by a quiescent gap before the main burst comes out. The emission properties of the precursor is not too different from the main burst \citep[e.g.,][]{Lazzati2005,Burlon2008,Hu2014}.  A power density spectrum (PDS) analysis does not reveal quasi-periodic oscillations, but rather show a featureless power law \citep{Beloborodov2000,Guidorzi2012}. However, there is evidence of the superposition of a ``fast'' variability component on top of a ``slow" variability component \citep{Gao2012}.

\subsection{Spectral properties}

\subsubsection{Spectral models and multiple components}

When enough photons are detected (large enough of GRB fluence), the GRB spectra are usually fit by a phenomenological model called the ``Band" function \citep{Band1993}. It is essentially a broken power law model with a smooth (exponential) transition between the two segments. The typical low-energy and high-energy photon indices are $\alpha \sim -1 \pm 1$ and $\beta \sim -2^{+1}_{-2}$ for the GRBs observed with BATSE in the 20-2000 keV band \citep{Preece2000}, and confirmed for GRBs detected by {\em Fermi}/GBM and Integral \citep{Zhangbb2011,Nava2011,Gruber2014}. The peak energy of a GRB (when $\beta<-2$ is satisfied) is defined by $E_p=(2+\alpha) E_0$, where $E_0$ is the break energy in the photon spectrum of the Band function. The distribution of $E_p$ is about several hundred keV for bright GRBs, but can vary from several keV for GRB 060218 \citep{Campana2006} to 15 MeV for GRB 110721A \citep{Axelsson2012}. The spectra of some GRBs may be fit with a cut-off power law, and some others with a single power-law function if the GRB is not bright enough or it is observed with a narrow band instrument such as {\em Swift} BAT, even though the true spectral shape may be Band-like \citep{Zhang2007a}.

In some GRBs, the superposition of multiple spectral components was observed. A thermal like spectral component claimed in the pre-{\em Fermi} era \citep{Ryde2005} was confirmed in the {\em Fermi} era. First, GRB 090902B was found to have a narrow Band component superposed on a power law component \citep{Abdo2009a}. The narrow Band component could be also fit as a multi-color blackbody \citep{Ryde2010}. When the time bin is small enough, the data could be even fit with a blackbody plus power law model \citep{Zhangbb2011}. A similar case may be made to the short GRB 090510 \citep{Ackermann2010}. Later, a weak thermal component was suggested to exist in the low-energy shoulder of the Band component of some GRBs \citep[e.g.,][]{Guiriec2011,Guiriec2013,Guiriec2015,Axelsson2012}. Some GRBs, on the other hand, show featureless Band-like spectra without the need of a thermal component \citep[e.g.,GRB 130606B,][]{Zhangbb2016}. Another spectral component is the power law component as seen in some GRBs, e.g. GRB 090902B and GRB 090510. This component extends several orders of magnitude from keV range sometimes all the way to the GeV range. It usually has a positive slope in $\nu F_\nu$ spectral representation, which means that there should be a high-energy $E_p$ in the GeV range. Indeed, GRB 090926 was found to have such a component with a high-energy cutoff \citep{Ackermann2011}. Some example GRB spectra are presented in Figure 2. Overall, GRB prompt emission spectra may have three distinct components \citep[][Fig.2d]{Zhangbb2011}: I. a Band component; II. a thermal component; and III. a power law component. Different spectral components may dominate in different GRBs. Indeed GRB spectral with different combinations of these three components have been observed. \cite{Guiriec2015} showed that all three components may exist in at least some bright GRBs.

\subsubsection{Spectral evolution}

Prompt emission spectra rapidly evolve with time. This can be seen from two different aspects. First, if one displays lightcurves in different energy bands, one would see significant ``spectral lags", i.e. a pulse observed in a softer band is usually broader than and lagged behind the same pulse observed in a harder band \citep{Norris2000,Liang2006b}. Second, if one displays time-dependent spectra, one would see that $E_p$ rapidly evolves with time. There are two types of evolution patterns \citep{Norris1986,Golenetskii1983,Lu2010,Lu2012}: hard-to-soft evolution and intensity tracking.

\subsection{Prompt emission in other wavelengths}

During the prompt emission phase, some GRBs are also detected in other energy bands. In the high-energy regime, GeV emission was detected in the prompt emission phase of several GRBs. In general, GeV emission is delayed with respect to the sub-MeV emission \citep[e.g.,][]{Abdo2009a,Abdo2009b}. Sometimes the GeV emission peak seems to align with some sub-MeV emission peaks, but in most cases, the GeV emission lasts longer than the MeV component \citep{Ghisellini2010,Zhangbb2011}, suggesting the emergence of a new emission component.

Prompt optical emission was detected in a few cases. Some GRBs show a tracking behavior between the optical and sub-MeV emission \citep[e.g., the famous ``naked-eye" GRB 080319B][]{Racusin2008}, while some others show a distinct peak near the end of prompt emission which is offset from the sub-MeV peaks \citep[e.g.,in GRB 990123][]{Akerlof1999}. The combination of these two components was also seen in some GRBs \citep[e.g. in GRB 050820A][]{Vestrand2006}.

On the other hand, extensive searches of high-energy neutrinos associated with GRBs during prompt emission phase have been carried out, but no detection has been made so far. An increasingly stringent upper limit of the prompt neutrino flux has been placed \citep{Abbasi2012,Aartsen2016}.

\subsection{Physical understanding}

No general consensus is reached in the community regarding how prompt emission properties are interpreted. This is because there are some uncertainties in the physical models of GRB prompt emission \citep[][for a review]{Zhangb2011}: jet composition (matter-dominated fireball or Poynting-flux-dominated outflow), energy dissipation mechanism and particle acceleration mechanism (internal shocks or magnetic reconnection), and radiation mechanism (synchrotron radiation or Comptonization of quasi-thermal photons). As a result, many models have been proposed to interpret GRB prompt emission. According to the location of emission site, these models may be grouped into three types: 1. the dissipative photosphere models \citep[e.g.,][;which interpret prompt emission as Comptonized quasi-thermal photons from the photosphere]{Rees2005,Peer2006,Giannios2007,Beloborodov2010,Lazzati2010,Toma2011,Murase2012}; 2. the internal shock models \citep[e.g.,][;which interpret prompt emission as synchrotron radiation of electrons accelerated in the internal shocks]{Rees1994,Daigne1998,Daigne2011}; and 3. the forced magnetic dissipation models with large emission radii \citep[e.g.,the internal collision-induced magnetic reconnection and turbulence, ICMART, model of][which interpret prompt emission as synchrotron radiation of electrons accelerated in the magnetic dissipation region]{Zhang2011}. These models have different magnetization parameter $\sigma$ in the emission region. The first two models have $\sigma \ll 1$, while the third model has $\sigma \geq 1$ in the emission region.\footnote{The so-called electromagnetic model \citep{Lyutikov2003} conjectures $\sigma \gg 1$ in the emission region which is close to the deceleration radius. It is hard to maintain such a huge $\sigma$ at the deceleration radius. The detections of the thermal components also disfavor this possibility.}

The erratic lightcurves of GRB prompt emission are interpreted differently in the three models. In the photosphere models, all the variabilities are directly produced by the generalized central engine (including the central black hole and the stellar envelope through which the jet propagates). The rapid variabilities may directly reflect the intrinsic fluctuation of the jet power from the central black hole, while the slow component may be caused by the modulation of the stellar envelope as the jet emerges from the central engine \citep{Morsony2010}. In the internal shock models, the observed emission is from many internal shocks, some at small radii (the fast component), and some at large radii (the slow component) \citep{Hascoet2012}. In the ICMART model, the slow component reflects the history of central engine activity. Each broad pulse is from one single emission unit (one ICMART event), which gives a seconds-long duration pulse as the emission region moves forward. The rapid variabilities may be caused by the mini-jets produced in the local reconnection regions within the jet \citep{Zhangzhang2014}.

It is agreed among the modelers that the thermal component observed in the spectra of some GRBs is emission from the outflow photosphere. The origin of the most common Band component is, however, subject to debate. One view is that it is also emission from the photosphere. The other view is that it is synchrotron emission from an optically thin region (internal shocks or ICMART regions). The simplest version of both models have trouble to interpret the typical low-energy photon index $\alpha \sim -1$: The photosphere model typically predicts $\alpha \sim +1.4$ \citep{Beloborodov2010,Deng2014}, whereas the synchrotron model typically predicts $\alpha \sim -1.5$ due to fast cooling of the electrons. In order achieve $\alpha \sim -1$, the photosphere model needs to introduce a special type of structured jet model \citep{Lundman2013}. Within the synchrotron model, $\alpha \sim -1$ (or an even harder spectrum) could be achieved if one considers the global decay of the magnetic field strength in the emission region, as is naturally expected in an expanding shell \citep{Uhm2014a}. The superposition of a thermal component on top of a Band component poses a challenge to the photosphere model\footnote{\cite{Vurm2011} interprets the Band component as the photosphere emission while the ``thermal'' bump as the synchrotron radiation.}. Within the internal shock and ICMART models, the Band component can be interpreted as the synchrotron radiation from an optically thin emission region (internal shocks or ICMART region), while the thermal component is the emission from the photosphere. Different bursts have different relative strength between the thermal and non-thermal components. This could be understood within the framework of a hybrid central engine \citep{Gao2015c} characterized by different dimensionless entropy $\eta$ (for the fireball component) and magnetization parameter $\sigma_0$ (the Poynting flux component).
The power law component extending from low energy to very high energy is not easy to explain. One possibility is that it actually includes two emission components, i.e. a low energy synchrotron component and a high-energy inverse Compton component. With certain parameters they may form an effective power law by coincidence \citep{Peer2012}.

Spectral evolution carries important clues regarding the origin of prompt emission. Spectral lags and $E_p$ evolution patterns all concern the broad pulses in GRB lightcurves rather than the rapid variabilities. This suggests that each broad pulse reflects emission from one single emitting shell which travels in space. \citep{Uhm2016a} showed that the properties of spectral lags can be reproduced within this picture given that the emission radius is large, the magnetic field strength in the emission region is decaying (as expected), and the emission region is undergoing bulk acceleration. The last requirement is consistent with dissipation of magnetic fields in a Poynting-flux-dominated outflow. The same theoretical framework can naturally explain $E_p$ hard-to-soft evolution \citep{Uhm2014a} and intensity tracking (Uhm and Zhang 2016b, in preparation). It is very difficult to interpret broad-pulse spectral lags and $E_p$ evolution patterns within the photosphere models and internal shock models. The stringent upper limits of neutrino flux during prompt emission phase is also consistent with the picture that most GRBs have a Poynting-flux-dominated jet composition \citep{Zhang2013,Aartsen2016}.

\section{Afterglow}
\label{sec:afterglow}

GRB afterglow is the broad-band emission detected after the prompt emission phase. It is characterized by a multi-segment broken power-law spectrum at a given instant, and a multi-segment broken power law decaying lightcurve in a given observational band. The afterglow was theoretically predicted \citep{Meszaros1997} before the first discoveries \citep{Costa1997,vanParadijs1997,Frail1997}. After two decades of observations, broad-band afterglows are well studied phenomenologically and physically.

\subsection{X-ray afterglow}

GRB X-ray afterglow was first discovered with BeppoSAX for GRB 970228 \citep{Costa1997}. A systematic study of X-ray afterglows started in the {\em Swift} era, when {\em Swift}/XRT regularly observe the GRB source starting from tens of seconds after the trigger. A canonical lightcurve can describe the X-ray afterglow lightcurves of most GRBs, which include five components \citep{Zhang2006,Nousek2006}: a steep decay phase, a shallow decay phase (or plateau), a normal decay phase, a late jet-break-like steepening, and erratic X-ray flares (Fig.\ref{fig:Xcanonical}). A small fraction of XRT lightcurves show a monotonic decay with a single power law \citep{Liang2009}. Empirically, \cite{OBrien2006} and \cite{Willingale2007} showed that the X-ray afterglow data can be fit by the superposition of a prompt emission component and an afterglow component, even though no physical model predicts the specific mathematical forms of the model. Some correlations between X-ray afterglow properties and prompt $\gamma$-ray properties have been found \citep{Grupe2013}.

The observational properties of the five afterglow components may be summarized as follows:
\begin{itemize}
\item Steep decay phase: The temporal decay slope of this segment is usually steeper than $-3$, sometimes even as steep as $-10$. This segment is usually smoothly connected to the last prompt emission pulse, suggesting that it is the tail emission of prompt emission \citep{Barthelmy2005}. Strong spectral softening is usually observed during the steep decay phase \citep{Zhangbb2007,Campana2006,Mangano2007,Butler2007}. The origin of the steep decay phase is usually attributed to the high-latitude ``curvature effect'', with a proper zero time adopted \citep{Kumar2000,Zhang2006,Liang2006a}. The rapid spectral evolution requires a curved spectrum at the end of prompt emission \citep{Zhangbb2007,Zhangbb2009}.
\item Shallow decay phase (plateau): This segment is usually seen in the early XRT lightcurves following the steep decay phase. Its decay slope ranges from 0 to $-0.5$ and the typical duration of this segment is $\sim10^4$ seconds with a fluence of $\sim 3\times 10^{-7}$ erg/s in the XRT band \citep{Liang2007a}. The typical photon index is $\sim 2.1$ \citep{OBrien2006,Liang2007a}, and there is no spectral evolution seen across the break from this segment to the normal segment afterwards \citep{Liang2007a}. The luminosity at the break time is anti-correlated with the break time in the burst frame \citep{Dainotti2010,Li2012}. In most cases, the shallow decay phase is followed by a normal decay with flux decreasing with time as $\sim t^{-1}$. These plateaus can be understood within the standard external shock model with continuous energy injection into the blastwave \citep{Zhang2006,Nousek2006,Rees1998,Dai1998a,Dai1998b,Zhang2011,Yu2007}. In a few cases, the shallow decay segment appears as a plateau followed by a very steep decay (with index around -9) in both long and short GRBs \citep{Troja2007,Liang2007a,Lyons2010,Rowlinson2010,Rowlinson2013}. These cases cannot be interpreted within the external shock models, and therefore are termed as ``internal plateaus''. They demand direct energy dissipation from a long-lasting central engine, likely a millisecond magnetar \citep{Troja2007,Lyons2010,Rowlinson2010,Rowlinson2013,Lyu2014a,Lyu2015}.
\item Normal decay and jet break: The normal decay phase has a typical decay index $\sim -1$, consistent with the prediction of the standard external forward shock model \citep{Meszaros1997,Sari1998}. In some GRBs, there is an additional steepening break at later times, after which the decay slope may reach $\sim -2$ \citep{Liang2008,Racusin2009}. This is consistent with the so-called jet break \citep{Rhoads1999,Sari1999c}. Some bursts were observed with {\em Swift}/XRT and/or {\em Chandra} weeks and even months after the GRB triggers, with no evidence of detecting a jet break in their X-ray lightcurves \citep[e.g.,][]{Grupe2006}.
\item X-ray flares: X-ray flares are detected in less than half of {\em Swift} GRBs. Their lightcurves show rapid rise and fall \citep{Burrows2005,Falcone2006,Romano2006} with strong spectral evolution \citep{Lazzati2007,Chincarini2007,Margutti2011}. The fluence of a flare is usually about $1\%-10\%$ of that of prompt emission, but occasionally may be comparable. Temporal and spectral analyses of X-ray flares reveal many properties analogous to prompt emission \citep{Burrows2005,Chincarini2010,Margutti2011}. They signify the reactivation of the central engine at later times \citep{Burrows2005,Fan2005,Zhang2006,Lazzati2007}.
\end{itemize}

\subsection{Optical Afterglow}
In the pre-{\em Swift} era, afterglow observations were mostly made in the optical bands after hours of GRB triggers. The late optical light curves usually show a single power law decay with a decay index of $\sim t^{-1}$. Occasionally, a two-segment broken power law from $\sim t^{-1}$ to $\sim t^{-2}$ was observed, with the temporal break defined as a ``jet break''. In some cases, an early optical flash was detected, which is characterized by a steep decay with $\sim t^{-2}$ \citep[e.g.,GRB 990123][]{Akerlof1999} and interpreted as emission from the reverse shock \citep{Meszaros1997,Meszaros1999,Sari1999a,Sari1999b}. \cite{Kann2010,Kann2011} presented optical afterglow lightcurves in both the observer frame and a common rest-frame at $z=1$ for a large sample of bursts (Fig. \ref{fig:OpticalLC3}a). Similar to the X-ray afterglow lightcurves, the optical afterglow lightcurves are also composed of several different emission components \citep[][Figure \ref{fig:OpticalLC3}b]{Li2012}: Ia. prompt optical flares; Ib. an early optical flare from the reverse shock; II. early shallow decay segment; III. the standard afterglow component (sometimes led by an afterglow onset hump due to deceleration); IV. the post jet break phase; V. optical flares; VI. rebrightening humps; VII. late supernova (SN) bumps. The components II-V can find their counterparts in the canonical X-ray lightcurve.

One interesting feature in some optical lightcurves is an early hump that marks the deceleration of the ejecta (e.g., \citealt[][for the thin shell case]{Rees1992,Meszaros1993,Sari1999b}, and \citealt[][for the thick shell case]{Kobayashi1999,Kobayashi2007}). The peak time in the lightcurve can be used to infer the initial Lorentz factor of the ejecta during the afterglow phase \citep[e.g.][]{Molinari2007,Xue2009,Melandri2010,Liang2010}. A statistical analysis \citep{Liang2013} suggests that the typical rising and decaying slopes of the onset hump are $\sim 1.5$ and $\sim -1.15$, respectively. The peak luminosity is anti-correlated with the peak time, $L_p\propto t_{p}^{-1.81\pm 0.32}$. Both $L_p$ and the isotropic energy release of the onset bumps are correlated with $E_{\gamma, \rm iso}$. \cite{Liang2010} discovered a correlation between $\Gamma_0$ and $E_{\gamma,iso}$ for a constant density medium. A similar relation between $\Gamma_0$ and $\gamma$-ray isotropic luminosity was also claimed \citep{Ghirlanda2011,Lyu2012}.

One important subject is the ``chromaticity" of the afterglows. Within the standard external forward shock models, optical and X-ray emission comes from the same synchrotron emission component. As a result, the hydrodynamical or geometrical temporal breaks (e.g. the energy injection break and jet break) in the lightcurves should be achromatic, i.e. a  break should occur simultaneously in the optical and X-ray bands. However, multi-wavelength observations suggested that some GRBs showed chromatic behaviors in the two bands \citep[e.g.,][Fig.\ref{fig:chromatic}]{Panaitescu2006,Fan2006b,Liang2007a,Liang2008}. A systematic study \citep{Wang2015} suggested that the majority of afterglows are consistent with being roughly achromatic, suggesting that the standard external shock models can account for most of the afterglow observations.

\subsection{Radio afterglow}

Compared with X-ray and optical afterglows, radio afterglow emission is more difficult to be detected with the available telescopes. About 30\% of GRBs were detected in radio, and this rate did not change before or after {\em Swift} was launched. Typically radio afterglow light curves at 8.5 GHz show initial rising and a peak time around 3-6 weeks in the rest frame \citep[Fig.\ref{fig:RadioLC},][]{Chandra2012}. This is consistent with the standard external forward shock model, with the peak corresponding to the crossing of the typical synchrotron frequency $\nu_m$ or the self-absorption frequency $\nu_a$ in the radio band. There is a clear relationship between the detectability of the radio afterglow and the fluence of a GRB \citep{Chandra2012}(Chandra \& Frail 2012). Some well-monitored bright GRBs show an early radio flare (e.g. GRB 990123, \citealt[][]{Kulkarni1999} and GRB 130427A, \citealt[][]{Anderson2014}), which is usually attributed to the emission from the reverse shock \citep{Sari1999a,Kobayashi2003a}.

\subsection{High energy afterglow}

Conventionally high-energy emission refer to photons above 100 MeV. The first detection of high-energy afterglow was in the Compton Gamma-Ray Observatory (CGRO) era, when GRB 940217 was detected in GeV energies hours after the burst was over \citep{Hurley1994}. {\em Fermi}/LAT detected a small fraction of GRBs in the GeV range. Most of these GeV-detected GRBs have their high-energy emission lasting longer than the prompt emission itself, suggesting an afterglow origin \citep[e.g.,][]{Abdo2009a,Abdo2009b,Ackermann2010,Ackermann2011}. Systematic studies suggest that the lightcurves typically show a power law decay, with the possibility of an early steep-to-shallow break transition \citep{Ghisellini2010,Zhangbb2011,Ackermann2013}.

\subsection{Physical understanding}

The standard theoretical framework to understand GRB afterglows is the external shock model. As the relativistic ejecta (also termed as fireball) is decelerated by the circumburst medium, either a constant density ISM or a stratified stellar wind, a pair of shocks are developed: one long-lasting forward shock propagating into the medium and a short-lived reverse shock propagating into the ejecta \citep{Meszaros1993,Sari1995}. Electrons are accelerated from both shocks, which radiate broad-band synchrotron radiation with characteristic frequencies evolving with time. The forward shock gives rise to the broad-band long-lasting afterglow \citep{Meszaros1997,Sari1998,Dai1998c,Chevalier2000,Huang2000} and a short-lived reverse shock may give additional significant optical and radio emission early on \citep{Meszaros1997,Meszaros1999,Sari1999a,Sari1999b,Kobayashi2000,Zhang2003,Kobayashi2003a,Kobayashi2003b,Wu2003,Fan2004,Zhang2005,Resmi2016}. For a comprehensive review of all the external shock models see \citet[][]{Gao2013a}. If the central engine is long lasting \citep{Dai1998a,Dai1998b,Zhang2001,Zhang2002} or if the ejecta has a stratified Lorentz factor distribution \citep{Rees1998,Sari2000}, continuous energy injection into the blastwave (defined as the region between the forward shock and reverse shock) is possible, so that the lightcurve decay slopes are shallower. In these cases, the reverse shock is long lived. If the reverse shock is more magnetized, these long-lasting reverse shocks may be the dominant emitter of afterglow emission \citep{Uhm2007,Genet2007,Uhm2012}.

Over the years, GRB afterglow data have been extensively modeled using the external shock model. In the pre-{\em Swift} era, afterglow observations were carried out hours after the bursts, and the data are in general well-interpreted by the models \citep[e.g.,][]{Panaitescu2001,Panaitescu2002,Yost2003}. {\em Swift} unveiled the early afterglow phase and discovered some unpredicted features (e.g. the steep decay phase, shallow decay phase, and X-ray flares). The theoretical framework in the post-{\em Swift} era invokes emission from both the external shock and internal energy dissipation regions (internal shocks or magnetic dissipation regions) due to the late central engine activities. Shortly after {\em Swift} started to detect early afterglow regularly, \cite{Zhang2006} suggested that the steep decay phase and the X-ray flares are of an internal origin, and that the shallow decay, normal decay and late steepening phases are from the external shock with the shallow decay phase being due to continuous energy injection into the blastwave. Such a simple paradigm remains valid to interpret most of GRB afterglows after a decade of observations.

In the optical band, the contamination of internal emission from late central engine activities is less, and most lightcurves can be interpreted within the standard forward shock model. Some bursts have early reverse shock component, but some others do not have evidence of reverse shock emission. The diversity may be attributed to different magnetization degree $\sigma$ of the ejecta. A moderately magnetized reverse shock ($\sigma < 0.1$) is favorable to produce a bright optical flash due to the enhanced synchrotron emission in the reverse shock region \citep{Fan2002,Zhang2003,Kumar2003}, but not necessarily a bright radio flash due to synchrotron self-absorption in the forward shock region \citep{Resmi2016}. An ejecta with $\sigma \ll 1$ (weak synchrotron emission) or $\sigma \geq 1$ (weak or no reverse shock) is not expected to produce bright reverse shock emission in optical \citep{Zhang2005}. The early optical afterglow data of a large sample of GRBs are consistent with a simple picture that GRB ejecta have a distribution of the magnetization parameter $\sigma$ among different bursts \citep{Gao2015b}.

The chromatic nature of afterglow observed in some GRBs cast doubts on interpreting all the afterglow data using the standard forward shock model. At least two different emission components are needed to interpret the data. One possibility is that long-lasting central engine activities are responsible to the entire observed X-ray afterglow emission \citep[e.g.,][]{Ghisellini2007,Kumar2008a,Kumar2008b}, while the optical emission is from the external forward shock. A second possibility is that a long-lasting reverse shock and the forward shock are contributing emission in different energy bands \citep{Uhm2012,Uhm2014b}. Another possibility is to invoke a two-component jet with each component dominating emission in one observational band \citep[e.g.,][]{dePasquale2009}. A recent systematic study \citep{Wang2015} suggested that the fraction of GRB afterglows that are consistent with being achromatic is at least 50\% and may be as high as 90\%. This suggests that these unconventional models may not be needed to interpret the majority of GRBs.

It is the consensus in the community \citep{Kumar2009,Kumar2010,Zhangbb2011,He2011,Liu2011,Maxham2011} that GeV afterglow after the prompt emission phase is of an external forward shock origin, while GeV emission during the prompt emission phase may have a contribution from the internal dissipation region. The radiation mechanism is believed to be dominated by synchrotron radiation, although synchrotron self-Compton may have been detected in GRB 130427A \citep{Fan2013,Liu2013,Tam2013,Ackermann2014}.

\section{Classification schemes}
\label{sec:classification}

Diverse GRBs with different observational properties have been observed. Classifying GRBs has been a challenging task, and many suggestions have been made. In general, these proposed classification schemes can be divided in two categories: phenomenological classification schemes and physical classification schemes.

\subsection{Phenomenological classification schemes}

The following classification schemes have been proposed in the literature:
\begin{itemize}
\item {\em Duration-hardness classification scheme:} This is the traditional long/soft vs. short/hard classification scheme \citep{Kouveliotou1993}. The main criterion is $T_{90}$, and hardness ratio is regarded as a supplementary criterion. The measured duration $T_{90}$ depends on the energy band and sensitivity of the detector \citep{Qin2013}, which may bring in ambiguity of classification in some bursts. For example, in the {\em Swift} era, some short GRBs are detected to be followed by soft extended emission lasting 10's to $\sim$ 100 s \citep{Norris2006}. Based on $T_{90}$ distribution information, some authors suggested a third, intermediate-duration class \citep[ee.g.,][]{Mukherjee1998,Horvath1998,Hakkila2003}. An ultra-long population was recently suggested \citep{Gendre2013,Levan2014}, even though whether it indeed form a distinct category is subject to debate \citep{Virgili2013,Zhangbb2014,Boer2015,Gao2015a}.
\item {\em High-luminosity (HL) vs. low-luminosity (LL):} Long GRBs are sometimes classified into two sub-categories based on their luminosities, with a separation line roughly at $(10^{48}-10^{49})~{\rm erg~s^{-1}}$. The LL-GRBs typically have long durations and smooth lightcurves, and have a much larger local event rate density than the traditional HL-GRBs \citep{Campana2006,Soderberg2006,Liang2007b,Virgili2009,Sun2015}.
\item {\em GRBs, X-ray rich GRBs, and X-ray flashes:} Based on spectral properties, long GRBs are sometimes further grouped into GRBs, X-ray rich GRBs, and X-ray flashes. The separation line is arbitrary since there is no clear bimodal distribution in the hardness distribution. Studies showed that these events form a continuum in their properties \citep[e.g.,][]{Sakamoto2005}. Most LL-GRBs are also X-ray flashes.
\item {\em Supplementary criteria:} Some prompt emission properties can be regarded as supplementary criteria for the duration classification scheme. Besides hardness ratio, spectral lags are often used to define whether a GRB is short \citep[short GRBs typically have negligible spectral lags, e.g.][]{Gehrels2006}. A combination of isotropic energy and rest-frame $E_p$ (the so-called $\epsilon=E_{\rm \gamma,iso}/E_{p,z}^{5/3}$ parameter) serves as a good separator between the long and short GRBs \citep{Lyu2010}. Also, the amplitude parameters ($f$ and $f_{\rm eff}$) can be very good criteria to distinguish between long and short \citep{Lyu2014b}.
\item {\em Afterglow-based classification schemes:} Based on optical afterglow data, GRBs can be classified as optically bright and optically dark GRBs \citep{Jakobsson2004,Rol2005}. Based on the X-ray afterglow data one may classify GRBs into those with X-ray flares and those without, or those following a canonical lightcurve and those following a single power law decay.
\end{itemize}

\subsection{Physical classification schemes}

The ultimate goal of studying astrophysical phenomenology is to uncover the physical nature of the sources. Based on observational properties and theoretical modeling of GRBs over the years, GRBs are physically classified into different categories.

\begin{itemize}
\item {\em Massive star (Type II) vs. Compact star (Type I) GRBs:} The two dominant categories of GRBs are the ones associated with deaths of massive stars and the ones not. The former category roughly correspond to the long duration GRBs, and latter roughly correspond to the short duration GRBs. However, counter examples do exist, e.g. SN-less long duration GRBs (GRB 060614 and GRB 060505) that are likely associated with compact stars \citep{Gehrels2006,GalYam2006,Fynbo2006,DellaValle2006}, and short GRBs that are likely associated with massive stars
    \citep[e.g. GRB 090426,][]{Levesque2010,Antonelli2009,Xin2011,Thone2011}. \cite{Zhang2007b} and \cite{Zhangb2006} proposed the Type I vs. Type II classification scheme in parallel with the short vs. long scheme, and \cite{Zhang2009} elaborated on the multiple observational criteria to identify the physical category of a certain GRB. Observationally, Type I and Type II GRBs show overlapping behaviors in all individual observational criteria \citep{Li2016a}, suggesting that multi-criteria are needed to perform classification. \cite{Zhang2009} proposed a flowchart to apply multi-criteria to classify GRBs, which was successfully applied to GRBs before 2011 \citep{Kann2010,Kann2011}.
\item {\em Successful vs. choked jets:} Within long (Type II) GRBs, bursts may be further classified into successful and choked jets. The former apply to the traditional HL-GRBs, while the latter may apply to some LL-GRBs such as GRB 060218 \citep[e.g.][]{Nakar2012,Bromberg2012}.
\item {\em Different progenitor stars within Type II GRBs:} For GRBs related to massive stars, there might exist different progenitor star systems. The leading type of progenitor star is Wolf-Rayet stars, rapidly rotating massive stars with hydrogen and helium envelope lost before core collapse \citep{Woosley1993,MacFadyen1999}. The association of long GRBs with Type Ic SNe (no H and He lines) is consistent with such a progenitor picture. Ultra-long GRBs, on the other hand, may call for a different progenitor star system with a larger size, e.g. blue supergiants \citep{Meszaros2001,Kashiyama2013,Greiner2015}, but a strong case of a new type of progenitor is not established. Besides single star collapsars \citep{Woosley2006}, a variety of binary progenitor star systems have been also discussed \citep{Fryer1999,Ruffini2016}. No smoking-gun signature has been identified to judge whether long GRB progenitors are in single or binary star systems.
\item {\em Different progenitor stars within Type I GRBs:} For GRBs not related to massive stars, the leading models invoke mergers of neutron star (NS) binaries \citep{Paczynski1986,Eichler1989,Paczynski1991,Narayan1992,Rosswog2003,Rezzolla2011}, either two NSs (NS-NS), or a NS and a black hole (NS-BH). Other types of progenitors \citep[e.g. accretion-induced-collapse, AIC, of NSs,][]{Qin1998,Dermer2006} are also possible. A merger-origin of short GRBs may be verified in the future when gravitational wave chirp signals are discovered to be associated with short GRBs. It is interesting to mention the putative short GRB signal detected with {\em Fermi}/GBM that lagged behind the first gravitational wave event GW 150914 by 0.4 s \citep{Connaughton2016}. If this is the case, BH-BH mergers may also make short GRBs, which may require unconventional mechanisms to produce short GRBs \citep[e.g.][]{Zhangb2016,Loeb2016,Perna2016,Li2016c,Murase2016,Yamazaki2016,Veres2016}.
\item {\em GRBs with different central engines:} For both long and short GRBs, two types of central engines have been widely discussed in the literature: hyper-accreting black holes \citep{Narayan1992,Woosley1993,Popham1999} and rapidly rotating magnetars \citep{Usov1992,Dai1998a,Dai1998b,Zhang2001,Metzger2011}. Signatures of magnetar spindown have been seen in both long and short GRBs \citep[][and references therein]{Lyu2014a,Lyu2015}. In particular, a possible magnetar central engine for short GRBs \citep{Dai2006,Gao2006,Fan2006a,Metzger2008}, if verified by future gravitational wave data, would have profound implications in constraining poorly known neutron star equation of state \citep{Lasky2014,Lyu2015,Gao2016a,Li2016b} and in searching for electromagnetic counterparts of gravitational wave events \citep{Zhangb2013,Gao2013b,Yu2013,Gao2016b}.
\item {\em Other bursts no longer defined as GRBs:} Historically several phenomena were initially confused as GRBs and later found to be other distinct phenomena that also emit bursts of $\gamma$-rays. An early example was the so-called soft-gamma-ray repeaters (SGRs), now known as bursts from Galactic magnetars. Some SGRBs emit giant flares characterized with a short hard spike followed by oscillating X-ray pulses \citep[e.g.][]{Palmer2005}. The short hard spikes of SGR giant flares can be detectable up to 80 Mpc, which may be confused as short GRBs. A systematic search suggested that these giant-flare-origin short GRBs may comprise less than 5\% of the short GRB population \citep{Tanvir2005}. One candidate of such events, GRB 051103, has been discovered \citep{Frederiks2007}. Another example was ``GRB 110328'' which was later identified as a tidal disruption event that launched a jet beaming towards earth \citep{Bloom2011,Burrows2011} and renamed as Sw 1644+57. Finally, it has been long proposed that evaporation of primordial black holes with mass $\sim 10^{17}$ g \citep{Hawking1975} may be able to produce ultra-short GRBs. However, no robust evidence for the existence of these bursts has been collected so far \citep[][and references therein]{Ukwatta2016}.
\end{itemize}

\section{Summary}

After nearly half century of observations, rich data have been accumulated for GRB prompt emission and afterglow. Our physical understanding of GRBs, including their progenitor systems, central engines, jet properties (composition, Lorentz factor), energy dissipation and radiation mechanisms, as well as the interaction between jet and ambient medium, has been greatly advanced. Yet, many open questions remain, which call for more dedicated multi-wavelength, multi-messenger observations that cover the full temporal and spectral windows in the prompt emission and afterglow phases.

\begin{acknowledgements} BZ acknowledges the NASA ADAP (NNX14AF85G) and ATP (NNX15AK85G) programs for supports. The research at Guangxi University was supported by the National Basic Research Program (973 Program) of China 2014CB845800, the National Natural Science Foundation of China (Grant No. 11533003,11603006,U1331202), One-Hundred-Talents Program of Guangxi colleges, Scientific
Research Foundation of GuangXi University (Grant No XGZ150299), and Guangxi Science Foundation (grant No. 2013GXNSFFA019001,2016GXNSFCB380005).
\end{acknowledgements}



\begin{figure*}
\centering
\includegraphics[width=0.7\textwidth]{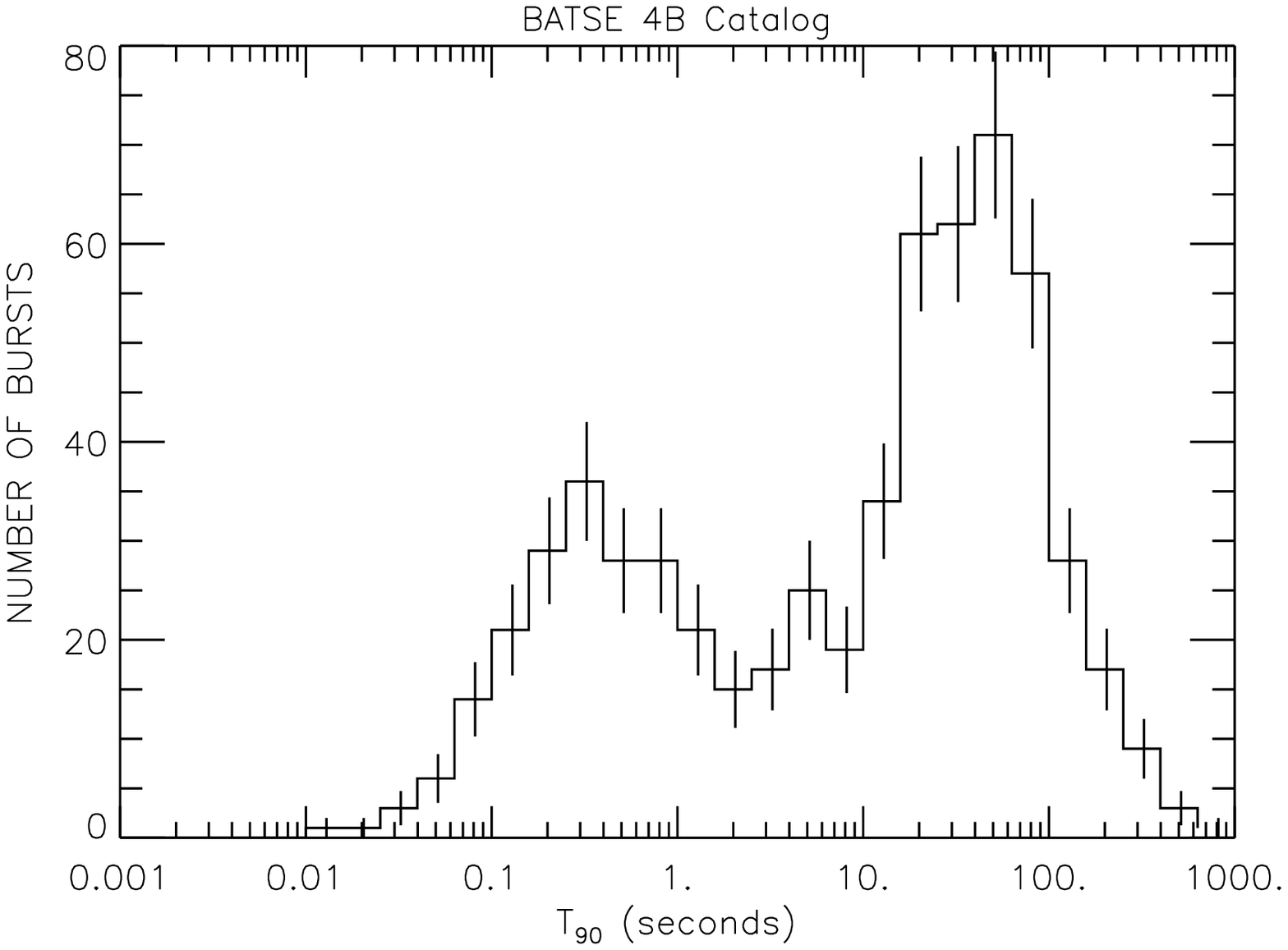}
\includegraphics[width=0.7\textwidth]{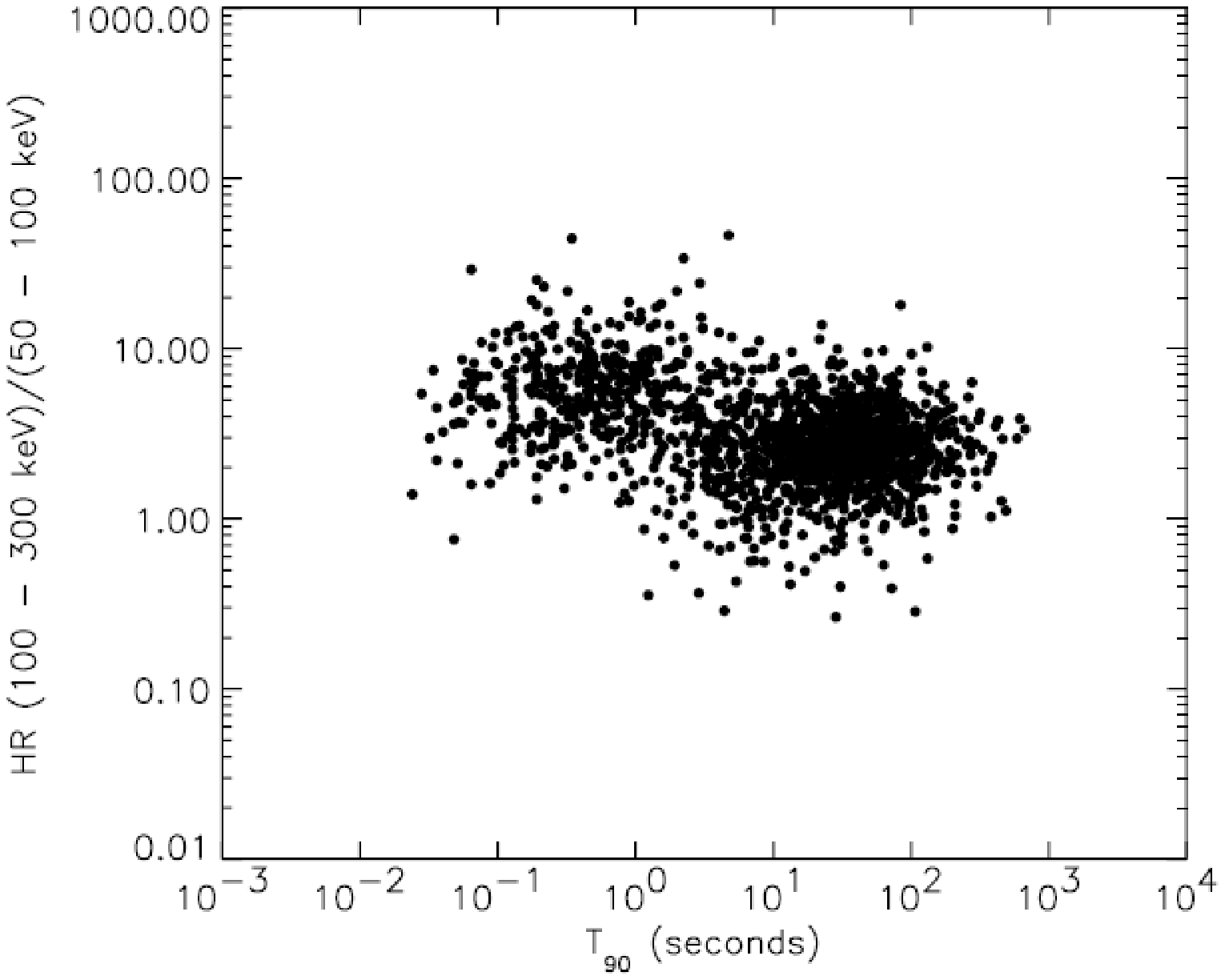}
\caption{{\em Left}: The bimodal distribution of GRB durations. {\em Right}: The two dimensional distribution of GRBs in the $T_{90}-{\rm HR}$ domain. From BATSE GRB catalogs (http://gammaray.msfc.nasa.gov/batse/grb/catalog/).}
\label{fig:T90}
\end{figure*}

\begin{figure*}
\centering
\includegraphics[width=0.4\textwidth]{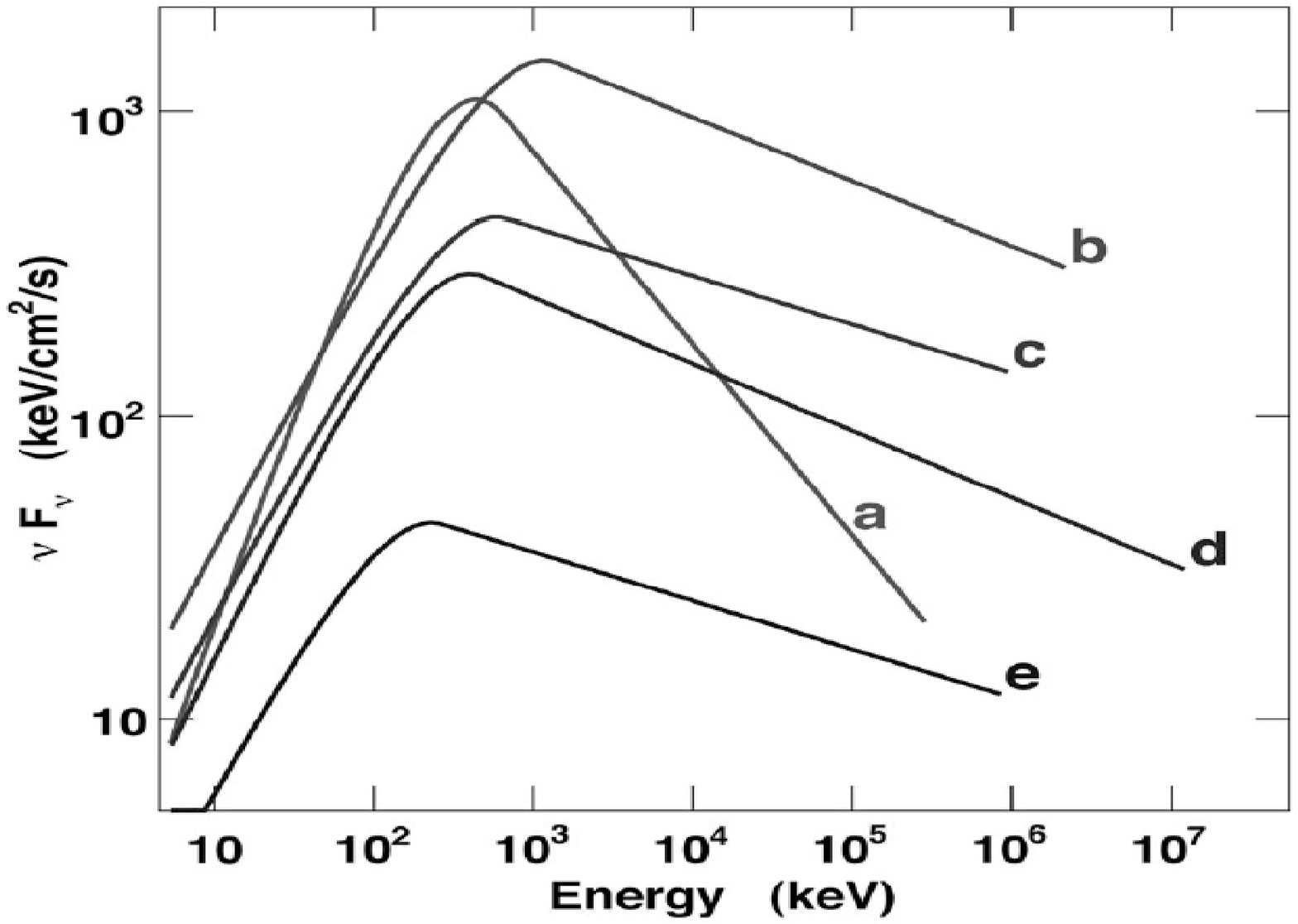}
\includegraphics[width=0.4\textwidth]{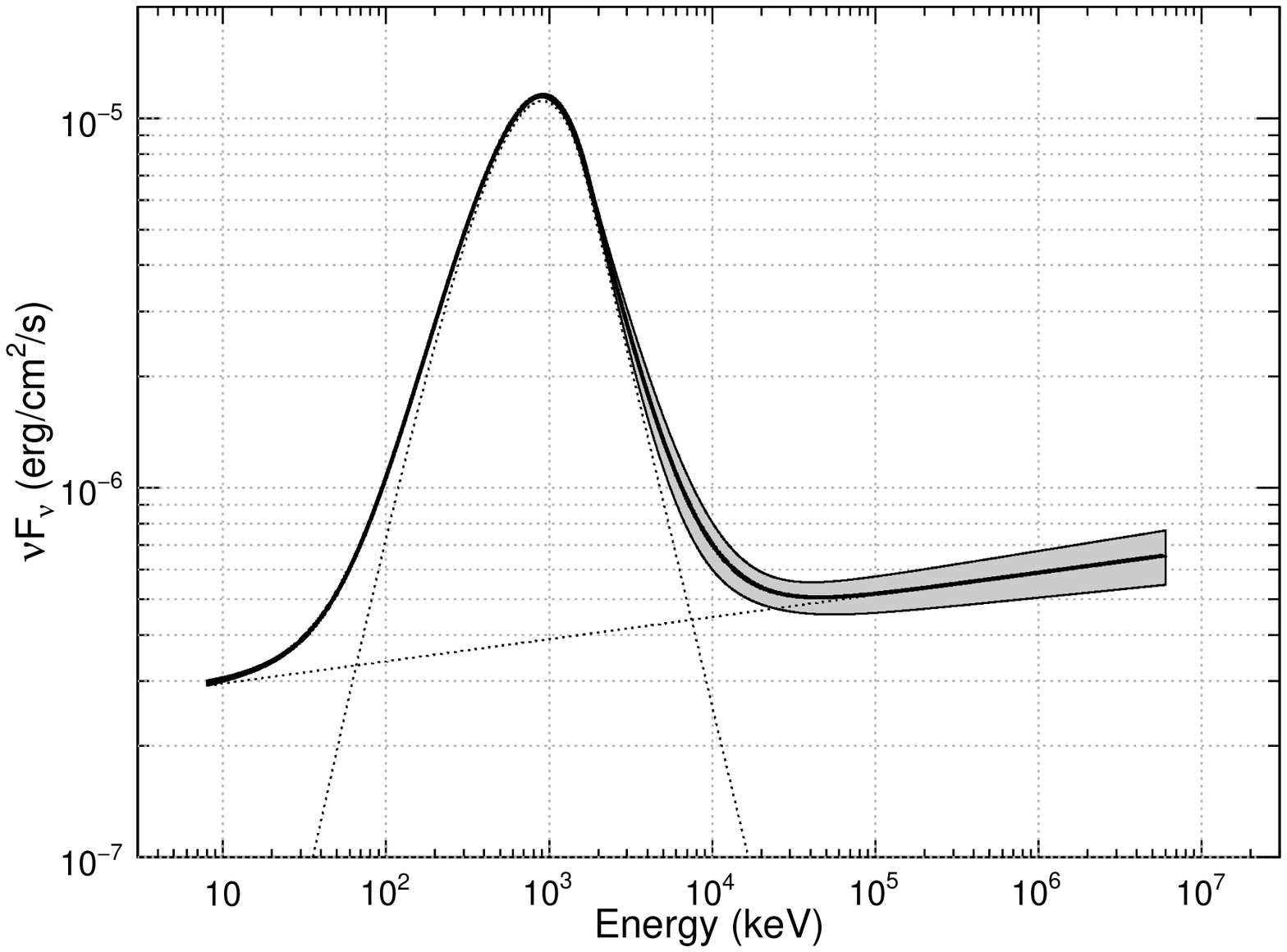}
\includegraphics[width=0.4\textwidth]{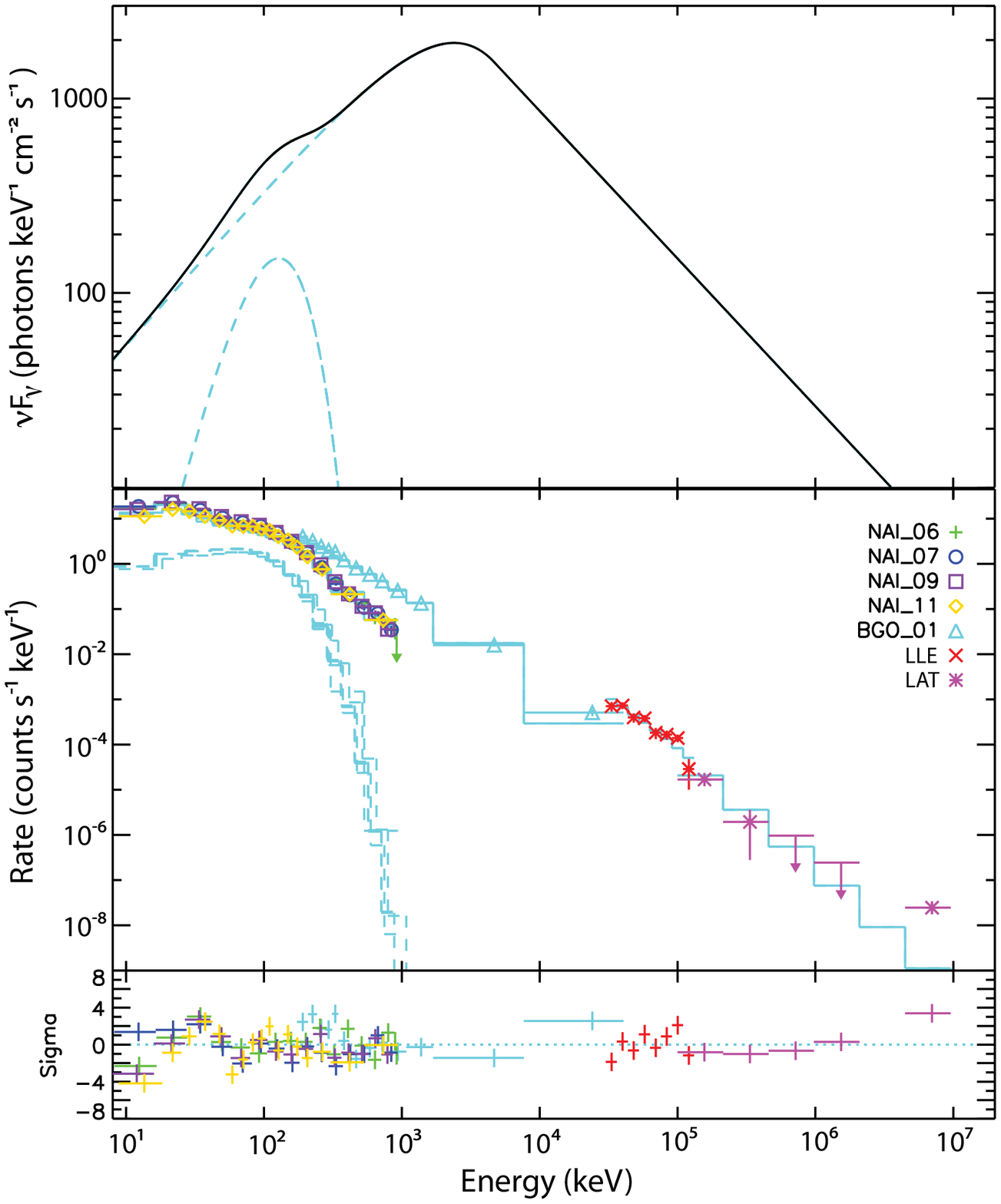}
\includegraphics[width=0.4\textwidth]{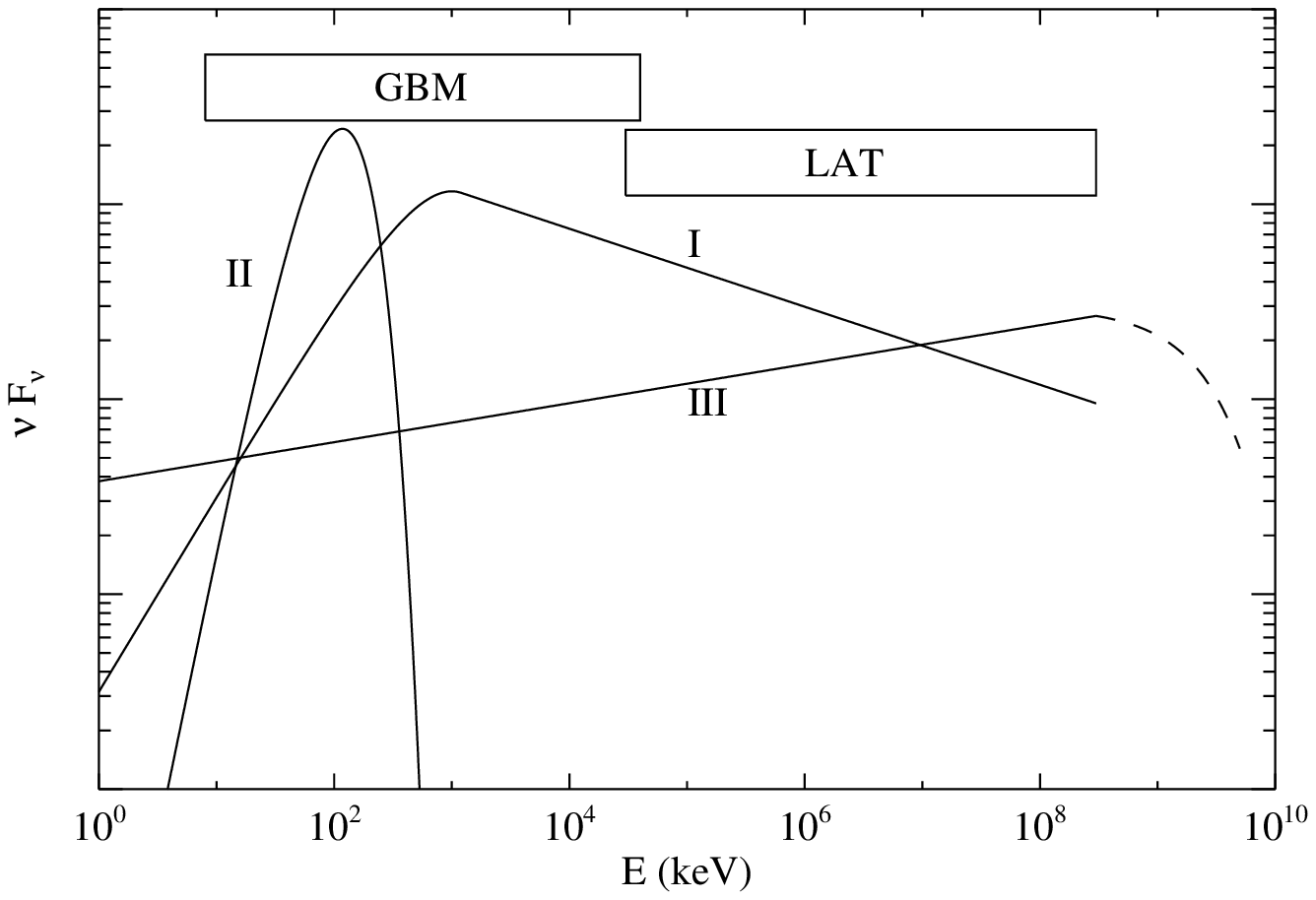}
\caption{Some example GRB spectra: (a) the Band function in GRB 080916c \citep{Abdo2009b}; (b) a quasi-thermal spectrum superposed on a power law component in GRB 090902B \citep{Abdo2009a}; (c) a thermal component superposed on a Band component in GRB 110721 \citep{Axelsson2012}; (d) three possible elemental spectral components that shape GRB spectra \citep{Zhangbb2011}.}
\label{fig:Spectrum}
\end{figure*}

\begin{figure*}
\centering
\includegraphics[width=0.6\textwidth,angle=-90]{f3a.eps}
\includegraphics[width=0.6\textwidth,angle=-90]{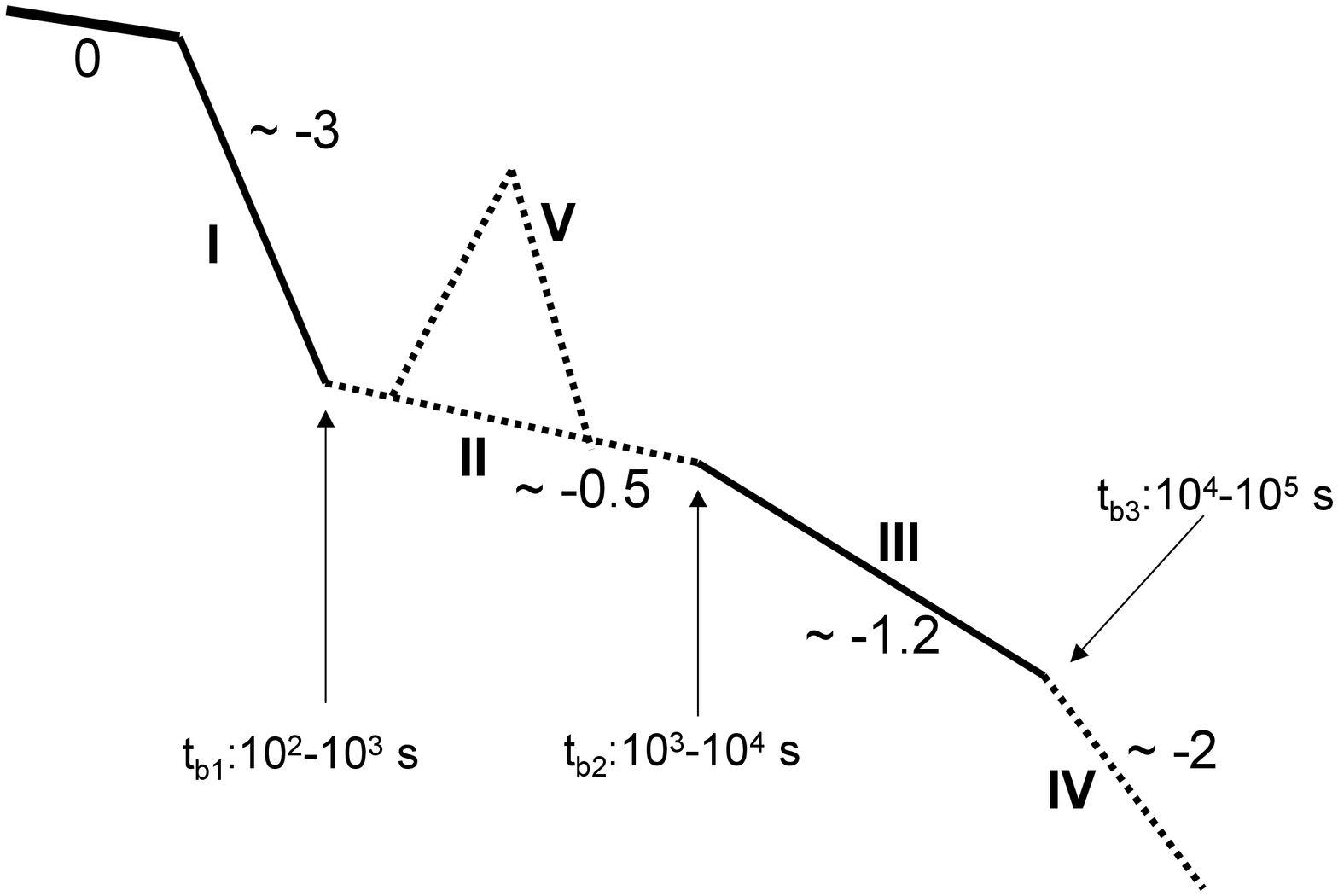}
\caption{{\em Left}: Examples of Observed XRT lightcurves
\citep{Nousek2006}. {\em Right}: A canonical XRT lightcurve \citep{Zhang2006}.}
\label{fig:Xcanonical}
\end{figure*}

\begin{figure*}
\centering
\includegraphics[width=0.7\textwidth]{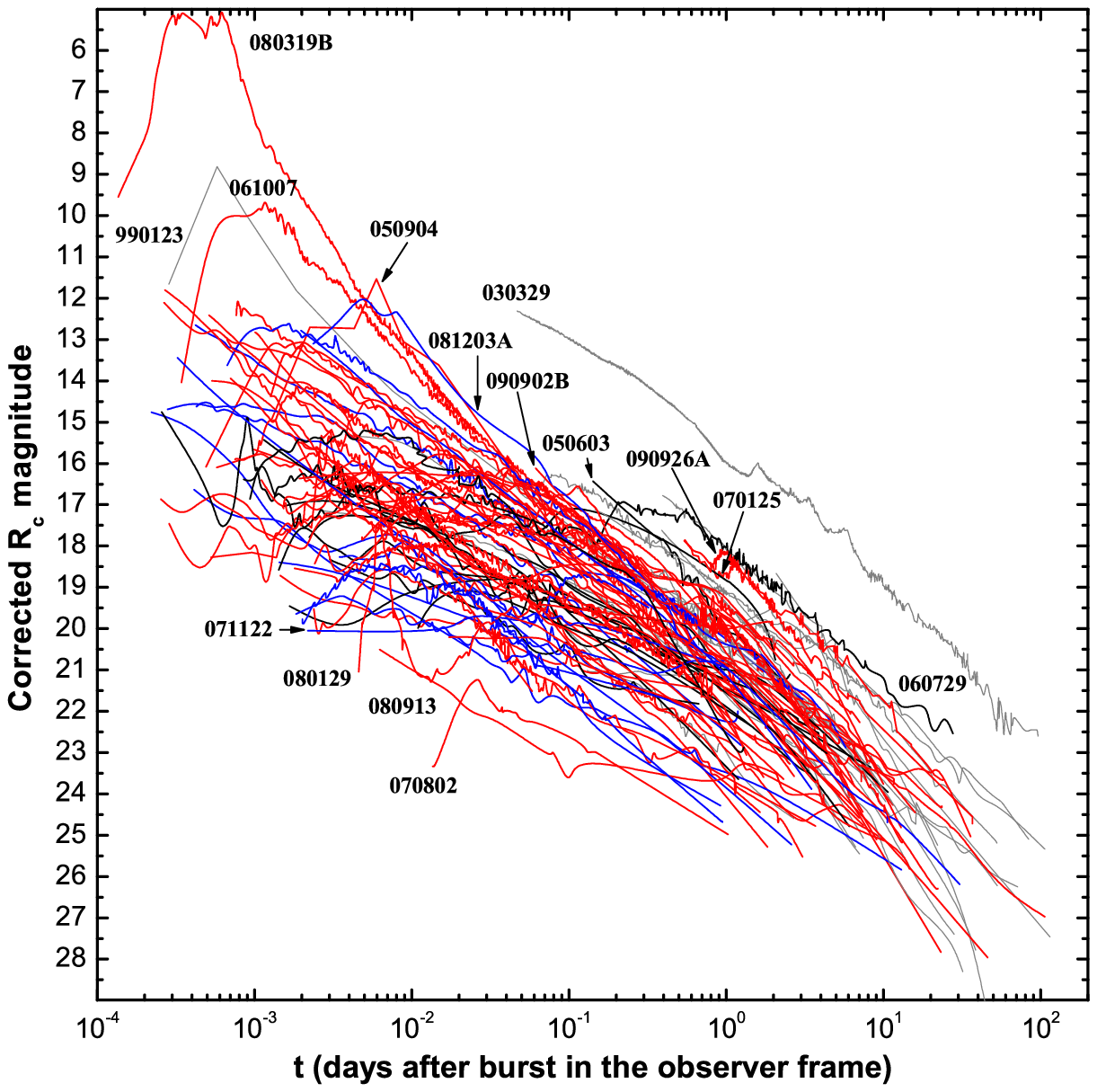}
\includegraphics[width=0.7\textwidth]{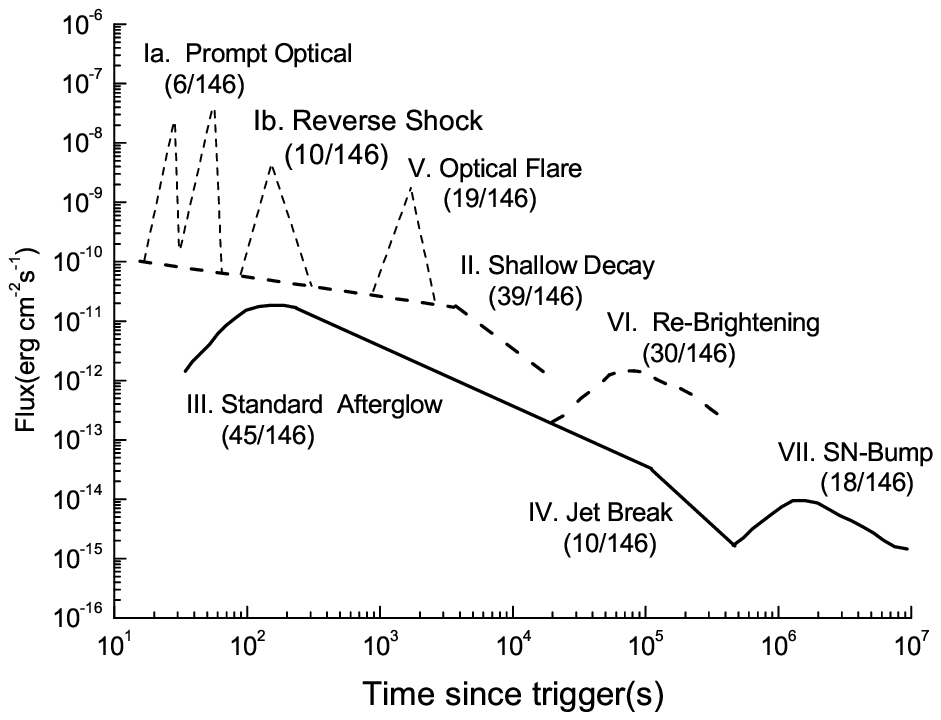}
\caption{(a) Observer-frame optical lightcurves for a sample of GRBs \citep{Kann2010};
(2) Synthetic schematic light curve of multiple optical emission
components, from \citep{Li2012}.}
\label{fig:OpticalLC3}
\end{figure*}

\begin{figure}
\centering
\includegraphics[angle=0, width=0.7\textwidth]{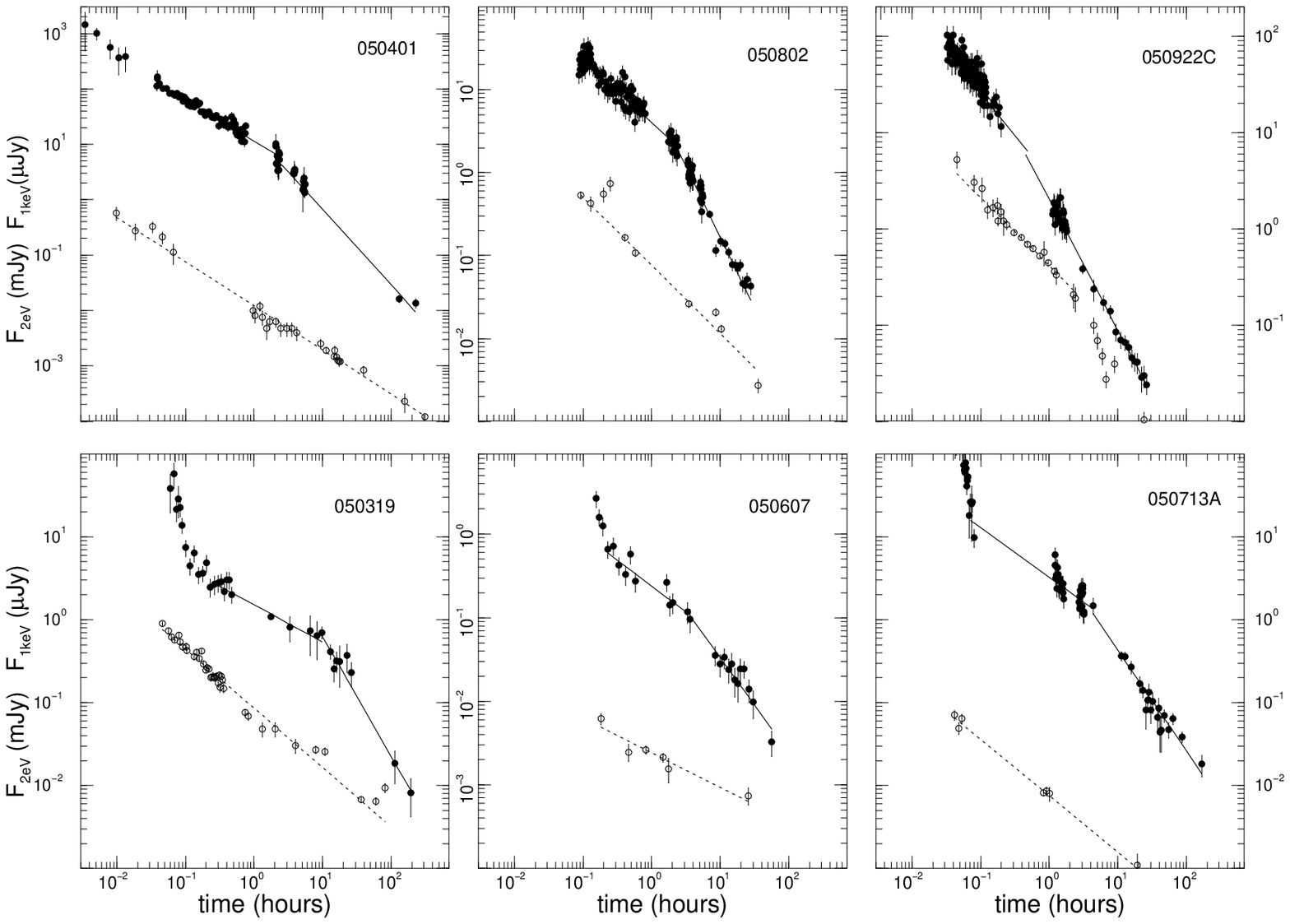}
\includegraphics[angle=0, width=0.7\textwidth]{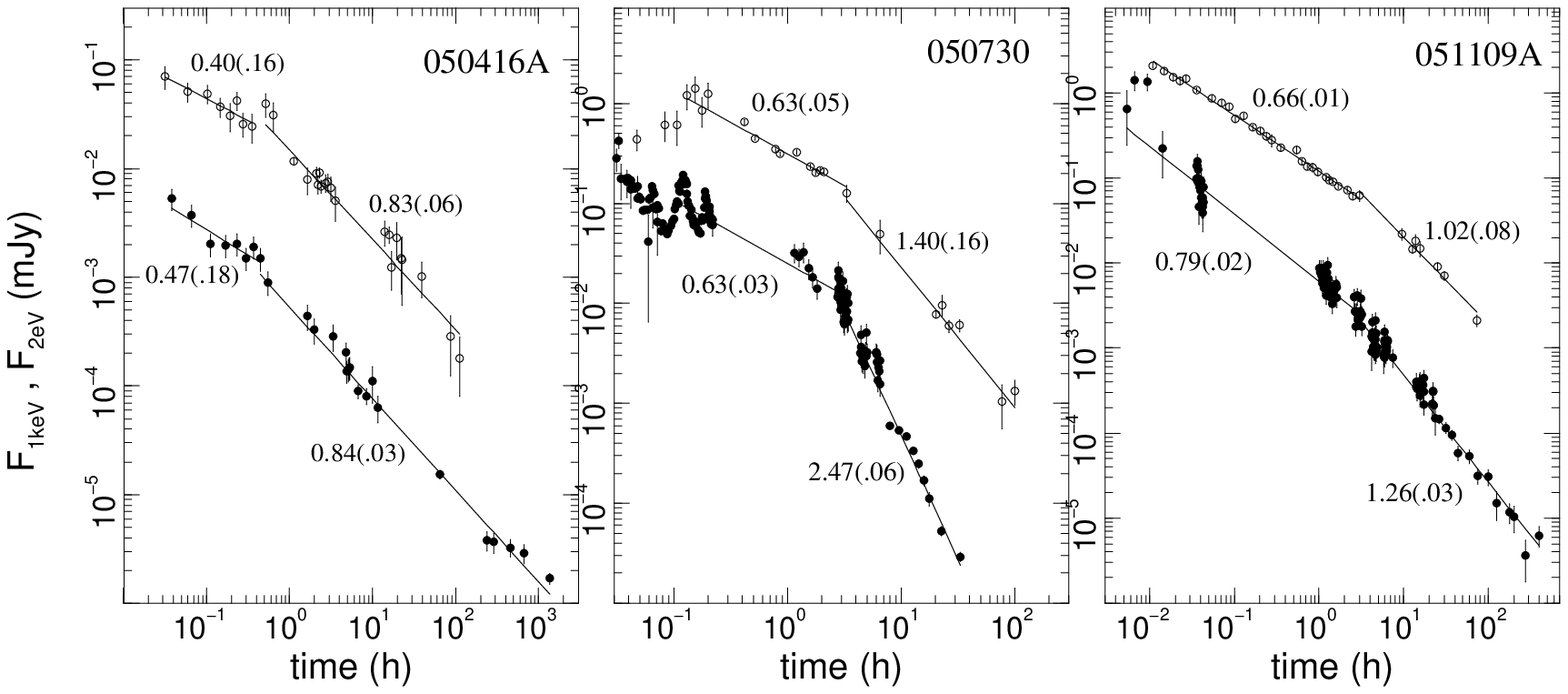}
\caption{Examples of chromatic and achromatic afterglow lightcurves, from \citep{Panaitescu2006}.}
\label{fig:chromatic}
\end{figure}

\begin{figure}
\centering
\includegraphics[angle=0, width=0.7\textwidth]{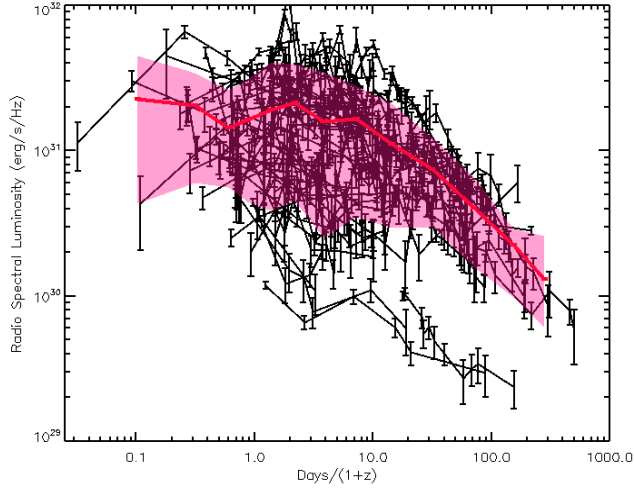}
\caption{The radio light curves at 8.5 GHz for the
long GRBs in the rest frame time. The red
solid line represents the mean light curve. The pink shaded area
represents the 75\% confidence band, from \cite{Chandra2012}.}
\label{fig:RadioLC}
\end{figure}

\clearpage




\end{document}